\documentclass[aps,preprint,a4paper]{revtex4-1}
\usepackage{graphicx}
\usepackage{amsmath}
\usepackage{amssymb}
\usepackage{bm}
\usepackage{amsthm}
\usepackage{bbold}
\usepackage{times}
\usepackage[T1]{fontenc}
\usepackage[utf8]{inputenc}
\usepackage{placeins}
\usepackage[version-1-compatibility]{siunitx}
\usepackage[pdfborder=0]{hyperref}
\hypersetup{
    colorlinks=true,
    linkcolor=blue,          
    citecolor=blue           
}
\theoremstyle{plain}
\newtheorem*{thm*}{Theorem}

\def\beq{\begin{equation}}
\def\eeq{\end{equation}}
\def\beqna{\begin{eqnarray}}
\def\eeqna{\end{eqnarray}}
\def\bea{\begin{array}}
\def\ea{\end{array}}

\def\etal{{\it et al }}
 
\begin{document}
\title{Amplification in parametrically-driven resonators near instability based
on Floquet theory and Green's functions}
\author{Adriano A. Batista}
\email{adriano@df.ufcg.edu.br}
\affiliation{
Departamento de Física\\
Universidade Federal de Campina Grande\\
Campina Grande-PB\\
CEP: 58109-970\\
Brazil}
\date{\today}
\begin{abstract}
Here we use Floquet theory to calculate the response of
parametrically-driven time-periodic systems near the onset of parametric 
instability to an added external ac signal or white noise.
We provide new estimates, based on the Green's function method,
for the response of the system in the frequency domain.
Furthermore, we present novel expressions for the power and 
noise spectral densities.
We validate our theoretical results by comparing our predictions for the
specific cases of a single degree of freedom parametric amplifier and of the
parametric amplifier coupled to a harmonic resonator with the
numerical integration results and with analytical approximate results obtained via
the averaging method up to second order.
\end{abstract}
\maketitle
\section{Introduction}
Amplification near bifurcation points has been studied since at
least the 80's with the publication of a seminal paper by Wiesenfeld and McNamara
on small ac signal amplification near period-doubling bifurcations
in nonlinear continuous dynamical systems \cite{wiesenfeld1985period}.
This work was later extended to account also for saddle-node,
transcritical, pitchfork, and Hopf bifurcations \cite{wiesenfeld1986small}.
In these systems, a small ac signal is added to a nonlinear
dynamical system that is on the verge of a local codimention-1 bifurcation. 
In their model a stable limit cycle solution of a nonlinear dynamical
system is perturbed by the addition of a small ac signal.
The response of the nonlinear system to this perturbation is 
equivalent to the dynamics of a parametric amplifier, i.e. of a linear parametrically driven system with an added ac drive.
Hence, Floquet theory can be used to calculate the gain of the amplification.
What they found is that the closer one is to the bifurcation
point the greater the response of the nonlinear system.
The complexity of the response is simplified by the fact that
the gain becomes dominated by the contribution of a single Floquet exponent.
This occurs because one of the Floquet exponents becomes zero  or crosses the
imaginary axis at the bifurcation point.
The generic analytical framework they developed to estimate this response is
based on this fact.
In addition to the response to a small added ac perturbation, Wiesenfeld
also analyzed the effect of noise near bifurcations in a paper with
Jeffries \cite{jeffries1985observation} and explained the effect of noisy
precursors of bifurcation in a periodically driven p-n junction.
In a subsequent paper, Ref.~\cite{wiesenfeld1985noisy}, he calculated estimates
for the noise spectral density (NSD).
Near the bifurcation point, he found that the NSD peak has a Lorentzian shape.

This concept of amplification near bifurcation points has proved to be very 
fruitful by spurring many applications and new ideas in diverse fields, such as Josephson
bifurcation amplifiers \cite{vijay2009invited}, bifurcation-topology amplifiers
\cite{karabalin2011signal}, tipping points in climate science
\cite{lenton2011early}, and ultra-sensitive nanoresonator charge detectors
\cite{dash2021ultra}.
To the author's knowledge, however, quantitative applications of the theory have
not been many.
In addition to that, many new experimental implementations of mechanical
parametric resonators and amplifiers have been developed recently, particularly
in nanomechanics
\cite{prasad2019gate,bothner2020cavity,lee2022giant,xu2022nanomechanical}.
Please see \cite{bachtold2022mesoscopic} for many more references on this topic.
In most of these works the resonators are pumped on the verge of the instability
threshold so as to increase gain.
This shows that there is an experimental need for a more quantitative model of
amplification near bifurcation points.

Here, we improve upon previous models so as to be able to quantitatively predict
the peak values of the power spectrum density (PSD) or the NSD, without the need
to rescale the Lorentzian lineshape to fit experimental data near the onset
of parametric instability. 
Furthermore, if one is not very close to the instability threshold, the Floquet
exponents could be a complex conjugate pair in the case of a
one-degree-of-freedom parametric amplifier as one can verify from the
Liouville's formula \cite{wiesenfeld1985virtual}. 
This fact results in split peaks such as predicted theoretically and observed
experimentally in Ref.~\cite{batista2022gain}. 
These split peaks have also been observed in parametric resonators with very
high quality factors \cite{miller2020spectral}. 
As it is, Wiesenfeld \etal theory cannot predict correctly these peak values and
likely they cannot be normalized to fit the data.
Here, we turn it more quantitative, with no need for rescaling the response to
an ac drive when the dynamics is known.
We point out that the main general assertions of their model remain intact, the
only improvement lies on the quantitative estimates for the gain of the
amplifier.

We believe that our theoretical contribution will make the generic model
of amplification near bifurcation points proposed by Wiesenfeld \etal more
accurate and, thus, likely to be even more useful and more widely used than it
presently is.
In addition to this, we provide an expression for the Green's
function that is not present in their papers or elsewhere to the author's
knowledge.
We already used approximate expressions for the time-domain Green's function of
the classical parametric resonator in Refs. \cite{batista2011signal, batista2012heating},
in which it was divided in two parts: a translationally invariant in time part
and another that is not translationally invariant in time.
We used this fact to obtain a more intuitive expression for the response
of the parametric resonator to an added ac signal or white noise. 
Here, we extended our Fourier analysis as presented in Ref.
\cite{batista2022gain}, based on the approximate averaging method, to the, in
principle, exact Floquet theory result. 
Furthermore, the results for the response of the resonator to the drive
and the NSD seem more intuitive physically than the ones provided in the literature.
The response of the resonator is similar to a combination of elastic (Rayleigh)
and inelastic (Raman) scattering processes.

This paper is organized as follows: In Sec.~\ref{theory}, we develop the
theory.
In subsection \ref{phi}, we present the basic facts about the
time-evolution operator or the fundamental matrix of linear dynamical systems.
In subsection \ref{amplification}, we present our results on parametric
amplification based on Floquet theory.
In subsection \ref{GF}, we show our main results on the generalized
Green's functions of time-periodic non-autonomous dynamical systems.
We also apply this method to obtain the PSD or the NSD when an external signal,
an ac drive or a white noise, is added to the unperturbed
parametrically-driven dynamical system. 
In Sec.~\ref{numerics}, we present and discuss our results.
In Sec.~\ref{conclusion}, we draw our conclusions.
In the Appendix, we present a proof of Floquet's theorem just for completeness
of this paper.






\section{Theory}
\label{theory}
\subsection{Preliminary: The Fundamental matrix}
\label{phi}
Given the linear non-autonomous ordinary differential equation (ODE) system
\beq
\dot x=A(t)x,
\label{ode}
\eeq
in which $x\in\mathbb{R}^N$ and $A(t)$ is a $N\times N$ real matrix.
The solution to this ODE system can be written as $x(t)=\Phi(t)x_0$, where
$x(0)=x_0$ and $\Phi(0)=\mathbb{1}$.
The $N\times N$ matrix $\Phi(t)$ obeys
\[
\dot \Phi(t)=A(t)\Phi(t).
\]
This time-evolution operator is known as the fundamental matrix.
It can be written as
\beq
\Phi(t)=\left[\bm{\phi_1(t)\;\phi_2(t)\;...\phi_N(t)}\right]
\label{funMatColumns}
\eeq
in which $\bm\phi_k$ is the $k$-th column.
With this notation $\Phi_{jk}=\bm{\phi}_{kj}$ is the $j$-th element of column
$\bm\phi_k$.
Since the time evolution of $x(t)$ is unique, $\Phi(t)$ is always non-singular
(i.e. invertible), therefore $\bm\phi_k$'s are linearly independent.

According to Floquet's theorem (see the Appendix for the proof) if
$A(t)$ is time periodic with period $T$,
the fundamental matrix can be written as $\Phi(t)=P(t)e^{Bt}$, where
$P(t+T)=P(t)$ is a periodic matrix.
The eigenvalues of $\Phi(T)=e^{BT}$ are called Floquet multipliers, while
the eigenvalues of $B$ are usually called Floquet exponents.
If $|v_\alpha\rangle$ is a column eigenvector of $B$ with eigenvalue $\rho_\alpha$,
then $e^{BT}|v_\alpha\rangle=e^{\rho_\alpha T} |v_\alpha\rangle$.
Note that the Floquet multipliers $e^{\rho_\alpha T}$ remain the same if $\rho_\alpha$ is replaced by
$\rho_\alpha+2i\pi n/T$.
To avoid uncertainty regarding this, we choose $\rho_\alpha$ to be in the first
Floquet zone, that is $-\pi/T<Im\rho_\alpha\leq\pi/T$.
Here, the eigenvectors $|v_\beta\rangle$ are columns and $\langle v_\alpha|$ are rows.
They are not just the transpose of one another,
rather, they are dual to each other, in such a way that 
$\langle v_\alpha|v_\beta\rangle=\delta_{\alpha,\beta}$.
In general, the eigenvectors of the matrix $B$ are not orthogonal, so the
relation of the vector basis, generated by $|v_\beta\rangle$'s, and the dual vector
basis, generated by $\langle v_\alpha|$, is similar to the relation between a lattice
basis vectors and its reciprocal lattice basis vectors.
In three dimensions, it is easy to construct the reciprocal lattice basis
vectors from the lattice basis vectors.
In higher dimensions, one can see that if the columns of a given matrix are the
eigenvectors $|v_\beta\rangle$, the dual basis vectors are the rows of the inverse
of this matrix. 
One should not confuse the columns of $\Phi(t)$ in
Eq.~\eqref{funMatColumns} with the eigenvectors $|v_\alpha\rangle$.

From linear algebra, one knows that the $|v_\alpha\rangle$'s form a complete basis
set, i.e. $\sum_\alpha|v_\alpha\rangle\langle v_\alpha|=\mathbb1$.
Therefore,
\beq
\begin{aligned}
\Phi(t) &=P(t)e^{Bt}=
\sum_\alpha|v_\alpha\rangle\langle v_\alpha|P(t)\sum_\beta e^{\rho_\beta
t}|v_\beta\rangle\langle v_\beta|
=
\sum_{\alpha, \beta}e^{\rho_\beta t}P_{\alpha\beta}(t)|v_\alpha\rangle
\langle v_\beta|,\\
\end{aligned}
\label{funMat}
\eeq
where $P_{\alpha\beta}(t)\doteq\langle v_\alpha|P(t)|v_\beta\rangle$.
Hence, one obtains 
$\Phi_{\alpha\beta}(t)\doteq\langle
v_\alpha|\Phi(t)|v_\beta\rangle=e^{\rho_\beta t}P_{\alpha\beta}(t)$
. 
On the other hand, we have 
\(
\Phi_{jk}=\langle j|\Phi(t)|k\rangle=
\sum_{\alpha, \beta}e^{\rho_\beta t}P_{\alpha\beta}(t)\langle j|v_\alpha\rangle
\langle v_\beta|k\rangle\)
, where
the canonical basis vector $|j\rangle$ is a vector with 1 at the $j$-th position
and zero elsewhere.
Similarly, we have
\beq
\begin{aligned}
\Phi^{-1}(t) &=e^{-Bt}P^{-1}(t)\doteq e^{-Bt}Q(t)
=\sum_{\alpha\beta}e^{-\rho_\alpha t}Q_{\alpha\beta}(t)|v_\alpha\rangle \langle
v_\beta|
\end{aligned}
\label{funMatInv}
\eeq
and 
\(
\Phi^{-1}_{\alpha\beta}(t)=\langle v_\alpha|\Phi^{-1}(t)|v_\beta\rangle
=e^{-\rho_\alpha t}Q_{\alpha\beta}(t).
\)
Notice that the columns of the fundamental matrix can be written as
\beq
\bm\phi_n (t)= \Phi(t)|n\rangle=
\sum_{\alpha, \beta}e^{\rho_\beta t}P_{\alpha\beta}(t)|v_\alpha\rangle
\langle v_\beta|n\rangle,
\eeq
which is more complicated than the  expression for $\bm\phi_k(t)$
given in references \cite{wiesenfeld1985noisy, wiesenfeld1986small}.

\subsection{Amplification near the onset of parametric instability}
\label{amplification}
The solution of the inhomogeneous real $n$-dimensional ODE system
\beq
\dot x=A(t)x+f(t),
\label{eq:paramp}
\eeq
can be related to the solution of the homogeneous system of Eq.~\eqref{ode}.
Applying the transformation, $x(t)=\Phi_0(t)y(t)$, where
$\dot\Phi_0(t)=A(t)\Phi_0(t)$, with initial conditions $\Phi_0(t_0)=\mathbb{1}$,
we obtain
\[
\Phi_0(t) \dot y= f(t).
\]
Hence, we find 
\[
y(t)=y(t_0)+\int_{t_0}^t\Phi_0^{-1}(t')f(t')dt'
\]
The solution $x(t)$ is thus given by
\beq
x(t)=\Phi_0(t)x(t_0)+\Phi_0(t)\int_{t_0}^t\Phi_0^{-1}(t')f(t')dt'
\eeq
If the quiescent solution of Eq.~\eqref{non-autonomous_system} is stable,
then for sufficiently large $t$, that is $t>>t_0$, $\Phi_0(t)x(t_0)\rightarrow 0$, and we can write
\beq
x(t)=\int_{t_0}^t\Phi_0(t)\Phi_0^{-1}(t')f(t')dt'
=\int_{t_0}^tP(t)e^{B(t-t')}P^{-1}(t')f(t')dt'
=\int_{t_0}^tP(t)e^{B(t-t')}Q(t')f(t')dt'.
\label{x_t}
\eeq
Hence, $G(t, t')=\Phi_0(t)\Phi_0^{-1}(t')=P(t)e^{B(t-t')}Q(t')$ is a
generalized Green's function.
The $n$-th component of the vector $x(t)$ is given by
\beq
\begin{aligned}
    x_n(t)&=\sum_m\int_{-\infty}^tG_{nm}(t, t')f_m(t')dt'\\ 
&= \sum_{j\alpha lm} \langle j|v_\alpha\rangle\langle v_\alpha|l\rangle 
P_{nj}(t)\int_{-\infty}^te^{\rho_\alpha 
(t-t')}Q_{lm}(t')f_m(t') dt',
\end{aligned}
\label{eq:x_n}
\eeq
where we chose $t_0=-\infty$.
With this choice of initial time we get rid of transients.
We point out that the expression in Eq.~\eqref{eq:x_n} is different from the
equivalent representation given in Ref.~\cite{wiesenfeld1986small}, specifically
in Eq.~(2.22) therein.
It seems there was a confusion between the basis of Floquet eigenvectors, i.e.
the eigenvectors of $B$ ($\{|v_\alpha\rangle|\; \alpha=1, 2,...,N\}$) and the canonical
basis ($\{|k\rangle|\;k=1, 2,...,N\}$).
The elements of the generalized time domain Green's function can be written as
\beq
G_{nm}(t, t')=\sum_{j\alpha l} \langle j|v_\alpha\rangle\langle v_\alpha|l\rangle
P_{nj}(t)e^{\rho_\alpha (t-t')} Q_{lm}(t').
\eeq
From the Floquet theorem, $P(t)=P(t+T)$ and $Q(t)=Q(t+T)$, where $T=2\pi/\omega$. 
Therefore, we can write the Fourier series of these periodic matrices as
\beq
\begin{aligned}
P(t)=\sum_{n=-\infty}^\infty p_{n}e^{in\omega t},\\
Q(t)=\sum_{n=-\infty}^\infty q_{n}e^{in\omega t}.
\end{aligned}
\label{eq:PQ}
\eeq
We notice that if the elements of the matrix $A(t)$ are real, then all the
elements of $\Phi(t)$ are also real.
Since $\Phi(T)=e^{BT}$, this implies that 
if $e^{BT}|v_\alpha\rangle=\mu_\alpha|v_\alpha\rangle$, then we also have 
$e^{BT}|v_\alpha\rangle^*=\mu_\alpha^*|v_\alpha\rangle^*$,
where the $^*$ symbol denotes complex conjugation.
Hence, this shows that if $|v_\alpha\rangle$ is a complex Floquet eigenvector
with Floquet eigenvalue $\mu_\alpha$, then $|v_\alpha\rangle^*$ is also
a Floquet eigenvector with eigenvalue $\mu_\alpha^*$.
Furthermore, we have $B|v_\alpha\rangle=\rho_\alpha|v_\alpha\rangle$, where
$\rho_\alpha=T^{-1}\ln\mu_\alpha$ and $B=\sum_\alpha\rho_\alpha|v_\alpha\rangle\langle v_\alpha|$ is a real matrix too.
Hence, $e^{-Bt}$ is real as well.
Therefore, $P(t)=\Phi(t)e^{-Bt}$ is real.
In addition to that, $Q(t)=e^{Bt}\Phi^{-1}(t)$ is also a real matrix.
Consequently, from Eq.~\eqref{eq:PQ}, we find that $p_{-n}=p_n^*$ and $q_{-n}=q_n^*$.

\subsubsection{A special case}
In what follows, we will focus on a special case of Eq.~\eqref{eq:paramp}, the
case in which $x(t)\in\mathbb R^2$ and $f_1(t)=0$ and 
$f_2(t)=F_0\cos(\omega_st)$.
We restrict our attention to the time evolution of
\beq
x_1(t) 
= \sum_{j\alpha l} \langle j|v_\alpha\rangle\langle v_\alpha|l\rangle 
P_{1j}(t)I_{\alpha l}(t),
\label{eq:x_1}
\eeq
where
\beq
I_{\alpha l}(t)=\int_{-\infty}^te^{\rho_\alpha(t-t')}Q_{l2}(t')f_2(t')dt'.
\label{eq:I_kl}
\eeq
Using the Fourier expansions of $P(t)$ e $Q(t)$ as written in Eq.~\eqref{eq:PQ}, we
can calculate the integral above and find analytical expressions for the matrix
elements $I_{\alpha l}(t)$, 
\beq
\begin{aligned}
    I_{\alpha l}(t)&=F_0\sum_{n=-\infty}^\infty q_{n,l2}\int_{-\infty}^te^{\rho_\alpha(t-t')}e^{in\omega
t'}\cos(\omega_st')dt'\\
&=F_0\sum_{n=-\infty}^\infty\frac{ q_{n,l2}e^{\rho_\alpha t}}2\left[
\dfrac{e^{[-\rho_\alpha+i((n+1)\omega+\Delta\omega)]t}}{-\rho_\alpha+i((n+1)\omega+\Delta\omega)}
-\dfrac{e^{-[\rho_\alpha+i((1-n)\omega+\Delta\omega)]t}}{\rho_\alpha+i((1-n)\omega+\Delta\omega)}
\right]\\
&=F_0\sum_{n=-\infty}^\infty\frac{q_{n,l2}}2\left[
\dfrac{e^{[i((n+1)\omega+\Delta\omega)]t}}{-\rho_\alpha+i((n+1)\omega+\Delta\omega)}
-\dfrac{e^{-i[(1-n)\omega+\Delta\omega]t}}{\rho_\alpha+i((1-n)\omega+\Delta\omega)}
\right],
\end{aligned}
\eeq
where $\Delta\omega =\omega_s-\omega$.
We are interested in studying the response near the onset of the parametric
instability.
This transition occurs when one of the Floquet exponents has zero real part as
can be gleaned from Eq.~\eqref{funMat}.
Below the threshold of instability all Floquet exponents have negative real
parts or are negative, while above it at least one Floquet exponent has positive real part or is
positive.
We obtain approximately
\beq
I_{\alpha l}(t)
\approx\frac{F_0}2\left[\dfrac{q_{1,l2}^*e^{i\Delta\omega t}}{-\rho_\alpha+i\Delta\omega}-
\dfrac{q_{1,l2}e^{-i\Delta\omega t}}{\rho_\alpha+i\Delta\omega}\right]
=-F_0Re\left\{\dfrac{q_{1,l2}e^{-i\Delta\omega t}}{\rho_\alpha+i\Delta\omega}\right\}.
\eeq
Hence, we find approximately
\beq
\begin{aligned}
x_1(t) 
&=-\frac{F_0}2\sum_{j\alpha l} \langle j|v_\alpha\rangle\langle v_\alpha|l\rangle 
P_{1j}(t)\left[\dfrac{q_{1,l2}^*e^{i\Delta\omega t}}{\rho_\alpha-i\Delta\omega}+
\dfrac{q_{1,l2}e^{-i\Delta\omega t}}{\rho_\alpha+i\Delta\omega}\right]\\
&\approx
r_s\cos[(\omega+\Delta\omega)t+\varphi_s]+r_i\cos[(\omega-\Delta\omega)t+\varphi_i]
\end{aligned}
\label{x_tFT}
\eeq
From Eq.~\eqref{eq:PQ}, 
we have $P_{1j}(t)\approx p_{1,1j}e^{i\omega t}+p_{-1,1j}e^{-i\omega t}$. 
Hence, we find the complex coefficients $a_s=r_se^{i\varphi_s}$ and 
$a_i=r_ie^{i\varphi_i}$ to be given by
\beq
\begin{aligned}
a_s =&-F_0\sum_{j\alpha l} \langle j|v_\alpha\rangle\langle
v_\alpha|l\rangle \frac{p_{1,1j}q^*_{1,l2}}{\rho_\alpha-i\Delta\omega}=
-F_0\langle1|p_{1}\sum_{\alpha} \frac{|v_\alpha\rangle\langle
v_\alpha|}{\rho_\alpha-i\Delta\omega}q^*_{1}|2\rangle
=-F_0\langle1|p_{1}\left(B-i\Delta\omega\right)^{-1}q^*_{1}|2\rangle, \\
a_i =&-F_0\sum_{j\alpha l} \langle j|v_\alpha\rangle\langle
v_\alpha|l\rangle \frac{p_{1,1j}q_{1,l2}}{\rho_\alpha+i\Delta\omega}
=-F_0\langle1|p_{1}\sum_{\alpha} \frac{|v_\alpha\rangle\langle v_\alpha|}{\rho_\alpha+i\Delta\omega}q_{1}|2\rangle
=-F_0\langle1|p_{1}\left(B+i\Delta\omega\right)^{-1}q_{1}|2\rangle. 
\end{aligned}
\label{signal_idler}
\eeq
From this, we can find the amplitudes of the signal and idler peaks.
Assuming that we are near (in parameter space) to the first parametric
instability region we have $\rho_2<<\rho_1\lesssim0$ (due to the Liouville's
formula). 
Consequently, the terms with Floquet
exponent $\rho_2$ decay much faster than the terms with exponent $\rho_1$.
The approximate amplitudes of the signal and idler peaks are, respectively,
\beq
\begin{aligned}
r_s =&F_0\left|\sum_{j\alpha l} \langle j|v_\alpha\rangle\langle
v_\alpha|l\rangle \frac{p_{1,1j}q^*_{1,l2}}{\rho_\alpha-i\Delta\omega}\right|
\approx\frac{F_0\left|\sum_{jl} \langle j|v_{1}\rangle\langle
v_{1}|l\rangle p_{1,1j} q^*_{1,l2}\right|}{\sqrt{\rho_1^2+\Delta\omega^2}}, \\
r_i =&F_0\left|\sum_{j\alpha l} \langle j|v_\alpha\rangle\langle
v_\alpha|l\rangle \frac{p_{1,1j}q_{1,l2}}{\rho_\alpha+i\Delta\omega}\right|
\approx\frac{F_0\left|\sum_{jl}\langle j|v_{1}\rangle\langle v_{1}|l\rangle
p_{1,1j}q_{1,l2}\right|}{\sqrt{\rho_1^2+\Delta\omega^2}}.
\end{aligned}
\eeq
The envelope is given by 
\beq
\begin{aligned}
    r(t) &=\pm\sqrt{u(t)^2+v(t)^2}, \mbox{ where}\\
    u(t) &= r_s\cos(\Delta\omega t+\varphi_s)+r_i\cos(\Delta\omega t+\varphi_i),\\
    v(t) &= r_s\sin(\Delta\omega t+\varphi_s)-r_i\sin(\Delta\omega t+\varphi_i).
\end{aligned}
\label{envelope}
\eeq
\subsection{The Green's function and the noise spectral density}
\label{GF}
In this subsection we develop an alternative method to obtain the response
of the non-autonomous system to an added signal that we believe is simpler
than the method we used above.
Using the Fourier series of Eq.~\eqref{eq:PQ}, we can write the generalized
retarded ($t>t'$) time-domain Green's function as
\beq
G(t, t') =P(t)e^{B(t-t')}Q(t')
=\sum_{n=-\infty}^\infty\sum_{m=-\infty}^\infty
p_ne^{B(t-t')}e^{i(nt+mt')\omega}q_m.
\eeq
We notice that it can be split in two parts
\beq
G(t, t') = G_0(t-t')+G_p(t, t'),
\eeq
a translationally invariant in time part, function of $t-t'$,  given by
\beq
G_0(t-t') =\sum_{n=-\infty}^\infty p_ne^{B(t-t')}e^{in\omega(t-t')}q_{-n}
=\sum_{n=-\infty}^\infty p_n\sum_\alpha|v_\alpha\rangle\langle v_\alpha|q_{-n}
e^{(\rho_\alpha+in\omega)(t-t')},
\eeq
and another part, which is a function of both $t$ and $t'$, given by
\beq
G_p(t, t') =\sum_{n=-\infty}^\infty
p_n\sum_{m=-\infty}^{\infty}{\!\!\!\!}^{'}e^{B(t-t')}e^{i\omega(nt+mt')}q_m
=\sum_{n=-\infty}^\infty
p_n\sum_{m=-\infty}^{\infty}{\!\!\!\!}^{'}\sum_\alpha|v_\alpha\rangle\langle
v_\alpha|q_me^{\rho_\alpha(t-t')+i\omega(nt+mt')},
\eeq
where $\sum'$ indicates that the summation omits $m=-n$.
The Fourier transform of $G_0(t-t')$ is
\beq
\begin{aligned}
\tilde G_0(\nu)&=\int_{-\infty}^\infty e^{i\nu\tau}G_0(\tau)d\tau=
\int_0^\infty e^{i\nu\tau}G_0(\tau)d\tau
=-\sum_{n=-\infty}^\infty \sum_\alpha\frac{p_n|v_\alpha\rangle\langle v_\alpha|q_{n}^*}{\rho_\alpha+i(\nu+n\omega)}\\
&=-\sum_{n=-\infty}^\infty p_n\left[B+i(\nu+n\omega)I\right]^{-1}q_{-n},
\end{aligned}
\label{G_0nu}
\eeq
where $\tau=t-t'$.
To lowest order approximation, it can be written  as
\beq
\begin{aligned}
\tilde G_0(\nu)&\approx
-\sum_\alpha\frac{p_1^*|v_\alpha\rangle\langle v_\alpha|q_1}{\rho_\alpha+i(\nu-\omega)}
-\sum_\alpha\frac{p_1|v_\alpha\rangle\langle v_\alpha|q_1^*}{\rho_\alpha+i(\nu+\omega)}\\
&=-p_1^*\left[B+i(\nu-\omega)I\right]^{-1}q_1-p_1\left[B+i(\nu+\omega)I\right]^{-1}q_1^*.
\end{aligned}
\eeq
Below, we can write the Fourier transform of $x(t)$ given in Eq.~\eqref{x_t}, with $t_0=-\infty$, as
\beq
\begin{aligned}
\tilde x(\nu)&=
\int_{-\infty}^\infty e^{i\nu t}\int_{-\infty}^tG_0(t-t')f(t')dt'dt
+\int_{-\infty}^\infty e^{i\nu t}\int_{-\infty}^tG_p(t,t')f(t')dt'dt\\
&=\tilde G_0(\nu)\tilde f(\nu)-\sum_{n=-\infty}^\infty
p_n\sum_{m=-\infty}^{\infty}{\!\!\!\!}^{'}
\left[B+i(\nu+n\omega)I\right]^{-1}q_m\tilde f(\nu+(n+m)\omega)\\
&=\tilde G_0(\nu)\tilde f(\nu)-\sum_{n=-\infty}^\infty
p_n\sum_{m=-\infty}^{\infty}{\!\!\!\!}^{'}\sum_\alpha|v_\alpha\rangle\langle v_\alpha|q_m\frac{\tilde f(\nu+(n+m)\omega)}{\rho_\alpha+i(\nu+n\omega)},
\end{aligned}
\label{fundEq1}
\eeq
where we used the relation
\beq
\int_{-\infty}^\infty e^{i\nu
t}\int_{-\infty}^te^{\rho_\alpha(t-t')+i\omega(nt+mt')}f(t')dt'dt =
-\frac{\tilde f(\nu+(n+m)\omega)}{\rho_\alpha+i(\nu+n\omega)}.
\eeq
The dynamical system response with lowest order corrections due to
parametric pumping is given by
\beq
\tilde x(\nu)\approx\tilde G_0(\nu)\tilde f(\nu)+G_+(\nu)\tilde f(\nu-2\omega)
+G_-(\nu)\tilde f(\nu+2\omega),
\label{fundEq2}
\eeq
where the coefficients are
\beq
\begin{aligned}
    G_+(\nu) &= -p_1^*\left[B+i(\nu-\omega)I\right]^{-1}q_1^*
    = -\sum_\alpha\frac{p_1^*|v_\alpha\rangle\langle v_\alpha|q_1^*}{\rho_\alpha+i(\nu-\omega)},\\
    G_-(\nu) &=-p_1 \left[B+i(\nu+\omega)I\right]^{-1}q_1
    =-\sum_\alpha\frac{p_1|v_\alpha\rangle\langle v_\alpha|q_1}{\rho_\alpha+i(\nu+\omega)}.
\end{aligned}
\label{ABcoefs}
\eeq
Note that, since $B$ is a real matrix, we have $G_+(-\nu)=G_-^*(\nu)$.
The expression given in Eq.~\eqref{fundEq1} is equivalent to the one obtained in
Ref.~\cite{batista2022gain}. 
While the previous expression is approximate, based on the 1st-order averaging method, and specific to the
one-degree-of-freedom parametric resonator with added noise, the
one proposed here is more general and exact, since it is a direct application 
of Floquet theory.
The expression given in Eq.~\eqref{fundEq2} is an approximation to
Eq.~\eqref{fundEq1} in which we throw out the higher harmonic terms.
Even this approximation, as we shall see, is considerably more accurate than the
equivalent averaging results.
It is noteworthy to mention that the matrix format of equations \eqref{ABcoefs}
is more compact, easier, and faster to deal numerically than the equivalent
ones with summations.
The  expressions in matrix format are better suited for modern programming
languages with builtin vectorization such as Python with the Numpy library.
Notice that the expression for $\tilde x(\nu)$ in Eq.~\eqref{fundEq1} with
$\tilde G_0(\nu)$ given by Eq.~\eqref{G_0nu} remain invariant when 
$\rho_\alpha$ is replaced by $\rho_\alpha+in\omega$.
On the other hand, the approximate expression given in Eq.~\eqref{fundEq2} is
not invariant under this transformation.
In this approximation, one has to choose $\rho_\alpha$ in the first Floquet zone.

It is worthwhile to explain to the reader how the matrix $B$ and its
eigenvectors and eigenvalues can be calculated numerically, since to the
author's knowledge this is not usually discussed in dynamical systems textbooks.
The general ODE system defined in Eq.~\eqref{ode} is integrated 
with the initial values set at $|k\rangle$, with $k=1,\dots N$ from
$t=0$ to $t=T$. 
The reason for this is that $\Phi(0)=\mathbb{1}$ is the $N\times N$ identity
matrix and $\Phi(t)$ obeys $\dot\Phi(t)=A(t)\Phi(t)$.
From these $N$ runs we compose $\Phi(T)$.
The numerical run with initial value $|k\rangle$ contributes the $k$-th column
of $\Phi(T)$. 
Based on the Floquet theorem, we find $\Phi(T)=P(T)e^{BT}=e^{BT}$. 
Then, we obtain numerically the eigenvalues $\mu_\alpha$ and
eigenvectors $|v_\alpha\rangle$ of $e^{BT}$. 
Note that $|v_\alpha\rangle$ are also eigenvectors of the matrix $B$. 
Due to fact that the set of all eigenvectors $|v_\alpha\rangle$
forms a complete set of the linear vector space of solutions of Eq.~\eqref{ode},
i.e. $\sum_\alpha|v_\alpha\rangle\langle v_\alpha|=\mathbb{1}$,
we find that 
$B=B\mathbb{1}=B\sum_\alpha|v_\alpha\rangle\langle v_\alpha|=\sum_\alpha\rho_\alpha |v_\alpha\rangle\langle v_\alpha|$.

\subsubsection{One-degree-of-freedom parametric amplifier}
For the case in which $N=2$ and the only nonzero term of $f(t)$ is $f_2(t)$,
the response $\tilde x_1(\nu)$ is given by
\beq
\tilde x_1(\nu)=\tilde G_{0,12}(\nu)\tilde f_2(\nu)
+G_{+,12}(\nu)\tilde f_2(\nu-2\omega)
+G_{-,12}(\nu)\tilde f_2(\nu+2\omega),
\eeq
where
$\tilde G_{0,12}(\nu)=\langle1|\tilde G_0(\nu)|2\rangle$,
$G_{+,12}(\nu)=\langle1|G_+(\nu)|2\rangle$, and
$G_{-,12}(\nu)=\langle1|G_-(\nu)|2\rangle$.
In the case in which $f_2(t)=F_0\cos(\omega_st+\varphi_0)$,
we obtain $\tilde f_2(\nu)=\pi
F_0\left[e^{i\varphi_0}\delta(\nu+\omega_s)+e^{-i\varphi_0}\delta(\nu-\omega_s)\right]$.
Hence, we obtain
\beq
\begin{aligned}
\tilde x_1(\nu)&=\pi F_0\bigg\{
\tilde
G_{0,12}(\nu)\left[e^{i\varphi_0}\delta(\nu+\omega_s)+e^{-i\varphi_0}\delta(\nu-\omega_s)\right]\\
&+G_{+,12}(\nu)\left[e^{i\varphi_0}\delta(\nu-\omega+\Delta\omega)+e^{-i\varphi_0}\delta(\nu-2\omega-\omega_s)\right]\\
&+G_{-,12}(\nu)\left[e^{i\varphi_0}\delta(\nu+2\omega+\omega_s)
+e^{-i\varphi_0}\delta(\nu+\omega-\Delta\omega)\right]\bigg\}.
\end{aligned}
\label{scattering}
\eeq
Applying the inverse Fourier transform to $\tilde x_1(\nu)$, we find
approximately
\beq
\begin{aligned}
    x_1(t)&=\frac{F_0}2\left[\tilde G_{0,12}(-\omega_s)e^{i(\omega_s t+\varphi_0)}
+\tilde G_{0,12}(\omega_s)e^{-i(\omega_s t+\varphi_0)}\right.\\
&\left.+G_{+,12}(\omega_i)e^{-i(\omega_it-\varphi_0)}
+G_{-,12}(-\omega_i)e^{i(\omega_it-\varphi_0)}\right]\\
&=\frac{F_0}2\left\{\left[\tilde G_{0,12}(\omega_s)e^{-i(\Delta\omega t+\varphi_0)}
+G_{+,12}(\omega_i)e^{i(\Delta\omega t+\varphi_0)}\right]e^{-i\omega t}
+c.c.\right\},
\end{aligned}
\eeq
where $\omega_i=\omega-\Delta\omega=2\omega-\omega_s$ is the idler angular frequency.
From the above equation, we obtain that the signal and idler amplitudes are
given by
\beq
r_s =F_0\left|\tilde G_{0,12}(\omega_s)\right| \mbox{ and }
r_i =F_0\left|G_{+,12}(\omega_i)\right|
\label{a_siGreen}
\eeq
and the envelopes of the cyclostationary time series $x_1(t)$ are given by
\beq
\pm F_0\left|\tilde G_{0,12}(\omega_s)e^{-i(\Delta\omega t+\varphi_0)}
+G_{+,12}(\omega_i)e^{i(\Delta\omega t+\varphi_0)}\right|.
\eeq
One can make the following analogy of the above process with scattering
\cite{huber2020spectral}.
The first two terms on the right side of Eq.~\eqref{scattering} correspond to
Rayleigh scattering, in which the scattering occurred without a change of
energy. 
The other terms correspond to Raman scattering where the incoming photon
scatters inelastically.
When $\omega_s<\omega$, the photon scatters and gains energy from the pump 
corresponding to a Stokes Raman scattering process.
When $\omega_s>\omega$, the photon scatters and loses energy to the pump
what corresponds to a anti-Stokes Raman scattering process.
Here we neglect the superharmonic terms.

The  PSD, see definition in Ref.~\cite{batista2022gain},  of $\tilde x_1$
obtained from Eq.~\eqref{scattering} when $\Delta\omega\neq0$ is given by
\beq
\begin{aligned}
    &S_{x_1}(\nu)=\lim_{\Delta\nu\rightarrow
0^+}\int^{\nu+\Delta\nu}_{\nu-\Delta\nu} 
\dfrac{\tilde x_1(-\nu)\tilde x_1(\nu')}{2\pi}\; d\nu'\\
&\approx\frac{\pi F_0^2}2\left\{\left|\tilde G_{0,12}(\nu)\right|^2\left[\delta(\nu+\omega_s)+\delta(\nu-\omega_s)\right]
+\left|G_{+,12}(\nu)\right|^2\delta(\nu-\omega_i)
+\left|G_{-,12}(\nu)\right|^2\delta(\nu+\omega_i)\right\}\\
&=\frac{\pi F_0^2}2\left\{\left|\tilde
G_{0,12}(\omega_s)\right|^2\left[\delta(\nu+\omega_s)+\delta(\nu-\omega_s)\right]
+\left|G_{+,12}(\omega_i)\right|^2\left[\delta(\nu-\omega_i)
+\delta(\nu+\omega_i)\right]\right\}.
\end{aligned}
\eeq
Notice that the integration in $\nu'$ eliminates one Dirac $\delta$ function 
and only pairs of $\delta$'s with the same argument contribute to the PSD,
since there is no overlap between $\delta$ functions with different arguments.
Also note that the statistical average is not necessary in the expression for
the PSD since $\tilde x_1$ is generated coherently.

When $\Delta\omega=0$, $\tilde x_1(\nu)$ is given by
\beq
\begin{aligned}
\tilde x_1(\nu)&\approx\pi F_0\left\{\left[
\tilde
G_{0,12}(\omega)e^{-i\varphi_0}+G_{+,12}(\omega)e^{i\varphi_0}\right]\delta(\nu-\omega)\right.\\
&\left.+\left[\tilde G_{0,12}^*(\omega)e^{i\varphi_0}+G_{+,12}^*(\omega)e^{-i\varphi_0}\right]\delta(\nu+\omega)
\right\}.
\end{aligned}
\eeq
From this, we find that the  PSD at degenerate parametric amplification 
can be written as
\beq
\begin{aligned}
    S_{x_1}(\nu)&=\lim_{\Delta\nu\rightarrow
0^+}\int^{\nu+\Delta\nu}_{\nu-\Delta\nu} 
\dfrac{\tilde x_1(-\nu)\tilde x_1(\nu')}{2\pi}\; d\nu'\\
&\approx\frac{\pi F_0^2}2\left|\tilde G_{0,12}(\omega)e^{-i\varphi_0}+G_{+,12}(\omega)e^{i\varphi_0}\right|^2\left\{\delta(\nu-\omega)
+\delta(\nu+\omega)
\right\}.
\end{aligned}
\eeq

In the case in which $\tilde f_2(\nu)=\tilde r(\nu)$, where $\tilde r(\nu)$ is
a white noise with zero mean and the statistical average
$\langle\tilde r(\nu)\tilde r(\nu')\rangle=4\pi D\delta(\nu+\nu')$,
where $D$ is the noise level.
We find that the NSD, as defined in Ref.
\cite{batista2022gain}, is given by
\beq
S_N(\nu)=\lim_{\Delta\nu\rightarrow 0^+}\int^{\nu+\Delta\nu}_{\nu-\Delta\nu} 
\dfrac{\langle \tilde x_1(-\nu)\tilde x_1(\nu')\rangle}{2\pi}\; d\nu'
=2D\left(\left|\tilde G_{0,12}(\nu)\right|^2+\left|G_{+,12}(\nu)\right|^2+\left|G_{-,12}(\nu)\right|^2\right).
\label{S_N}
\eeq
\section{Numerical results and discussion}
\label{numerics}
Here we apply the general theory developed in the previous section to two
specific dynamical systems.
The first one is a single parametric amplifier, whose equation of motion is
given by
\beq
\ddot x= -x-\gamma\dot x+F_p\cos(2\omega t)x+F_0\cos(\omega_s t+\phi_0).
\label{paramp}
\eeq
For this case the matrix $A(t)$ is given by
\[
A(t)=
\left[
\bea{cc}
0 & 1\\
-1+F_p\cos(2\omega t)&-\gamma
\ea
\right]
\]
and $f(t)$ is 
\[
f(t)=
\left[
\bea{c}
0\\
F_0\cos(\omega_s t+\phi_0)
\ea
\right].
\]
The second system is a parametric amplifier coupled to a harmonic resonator
with the following equations of motion
\beq
\begin{aligned}
\ddot x &=-\gamma_1 \dot x-[1-F_p\cos(2\omega t)]x-\beta_1 y+F_0\cos(\omega_s t+\phi_0),\\
\ddot y &= -\gamma_2\dot y-\omega_2^2y-\beta_2 x.
\end{aligned}
\label{2ModeParamp}
\eeq
This model is the linearized version of the model proposed by Singh \etal 
\cite{singh2020giant} to study frequency-comb spectrum generation
in parametrically-driven coupled mode resonators.
For this case the matrix $A(t)$ is given by
\[
A(t)=
\left[
\bea{cccc}
0 & 1 &0 &0\\
-1+F_p\cos(2\omega t)&-\gamma_1& -\beta_1 &0\\
0 &0 &0 &1\\
-\beta_2 &0 &-\omega_2^2 &-\gamma_2
\ea
\right]
\]
and $f(t)$ is 
\[
f(t)=
\left[
\bea{c}
0\\
F_0\cos(\omega_s t+\phi_0)\\
0\\
0
\ea
\right].
\]

\begin{figure}[!ht]
\centerline{\includegraphics[{scale=0.75}]{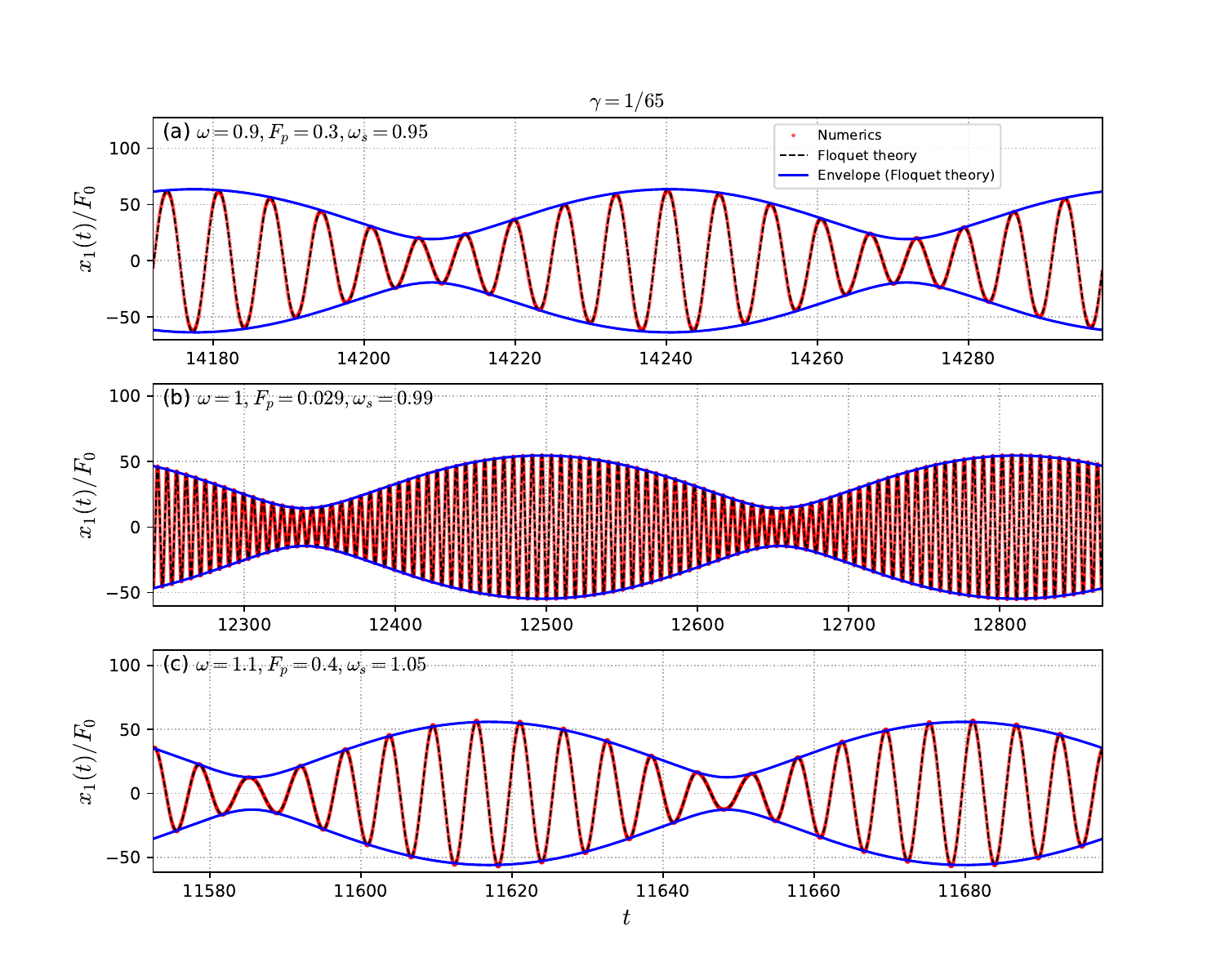}}
\caption{Single parametric amplifier. 
Comparison between a time series obtained from the numerical integration of Eq.~\eqref{paramp} and
the Floquet theory time-series fit from Eq.~\eqref{x_tFT} with corresponding
envelopes given by Eq.~(\ref{envelope}).  
The amplification here is set in quasi-degenerate mode with all parameters
indicated in the figure.}
\label{fig:envelopes}
\end{figure}

In Fig. \ref{fig:envelopes}, we show cyclostationary time series of the
numerical integration of Eq.~\eqref{paramp} compared with the predictions
of Floquet theory given by the first term of Eq.~\eqref{x_tFT}, while the envelopes are given
by Eq.~(\ref{envelope}).
In these examples only the harmonics with $n=\pm 1$ are enough to give a very
accurate description of the dynamics.
We needed contributions from both Floquet exponent terms to reach this
very high agreement, especially in the results of frames (a) and (c) which
present larger detunings and are not as close to the instability threshold.

\begin{figure}[!ht]
    \centerline{\includegraphics[{scale=0.75}]{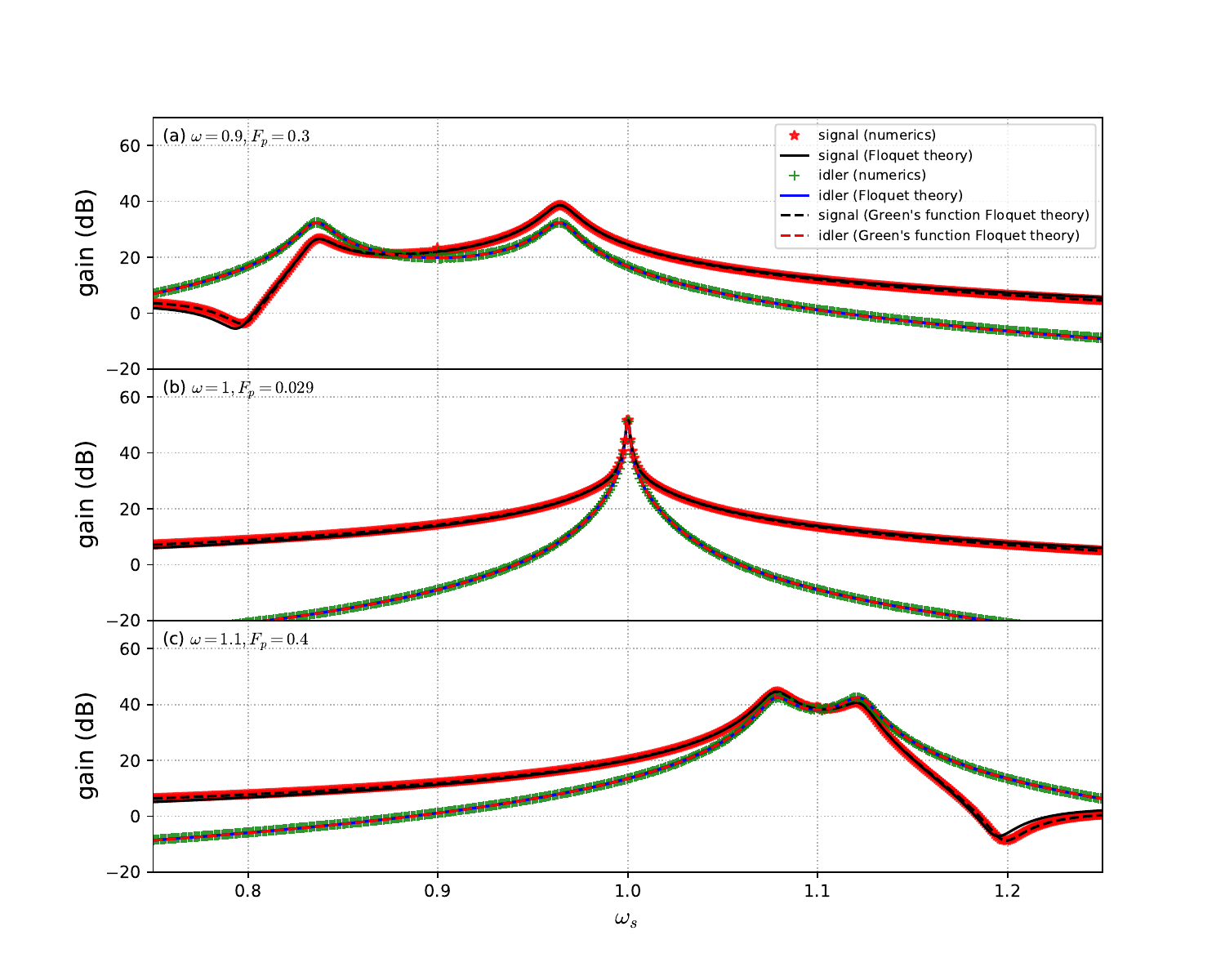}}
\caption{Floquet theory and numerical signal and idler response gains as a
function of the ac drive angular frequency for the
single parametric amplifier model defined in Eq.~\eqref{paramp}. 
The signal gain in dB scale is given by $20\log_{10}(r_s/F_0)$
and the idler gain in dB is given by $20\log_{10}(r_i/F_0)$.
The amplitudes $r_s$ and $r_i$ obtained from Eq.~\eqref{signal_idler} 
are represented by the full lines, while $r_s$ and $r_i$ obtained from
Eq.~\eqref{a_siGreen} are shown by the dashed lines.
In frames a) and c) the Floquet exponents are
complex, what causes the split peaks in the signal and idler responses, whereas
in frame b) they are real.
In all frames $\gamma=1/65$.}
\label{fig:signal_idler}
\end{figure}

In Fig. \ref{fig:signal_idler}, we show Floquet theory and numerical results
of the signal and idler gains, based on Eq.~\eqref{signal_idler}, as a function of $\omega_s$ for the single
parametric amplifier of Eq.~\eqref{paramp}.
We point out that only the lowest terms of the Floquet theory expansion are
enough to give a very accurate representation of the response of the parametric
amplifier.
There are only two slight disagreements on the far left side of frame (a) and on
the far right side of frame (c) for the signal response.
Note that this discrepancy disappears when we take into account the
off-resonance terms in the Floquet theory expansion for $a_s$ and $a_i$ using
the Green's method given in Eq.~\eqref{a_siGreen}.

\begin{figure}[!ht]
    \centerline{\includegraphics[{scale=0.75}]{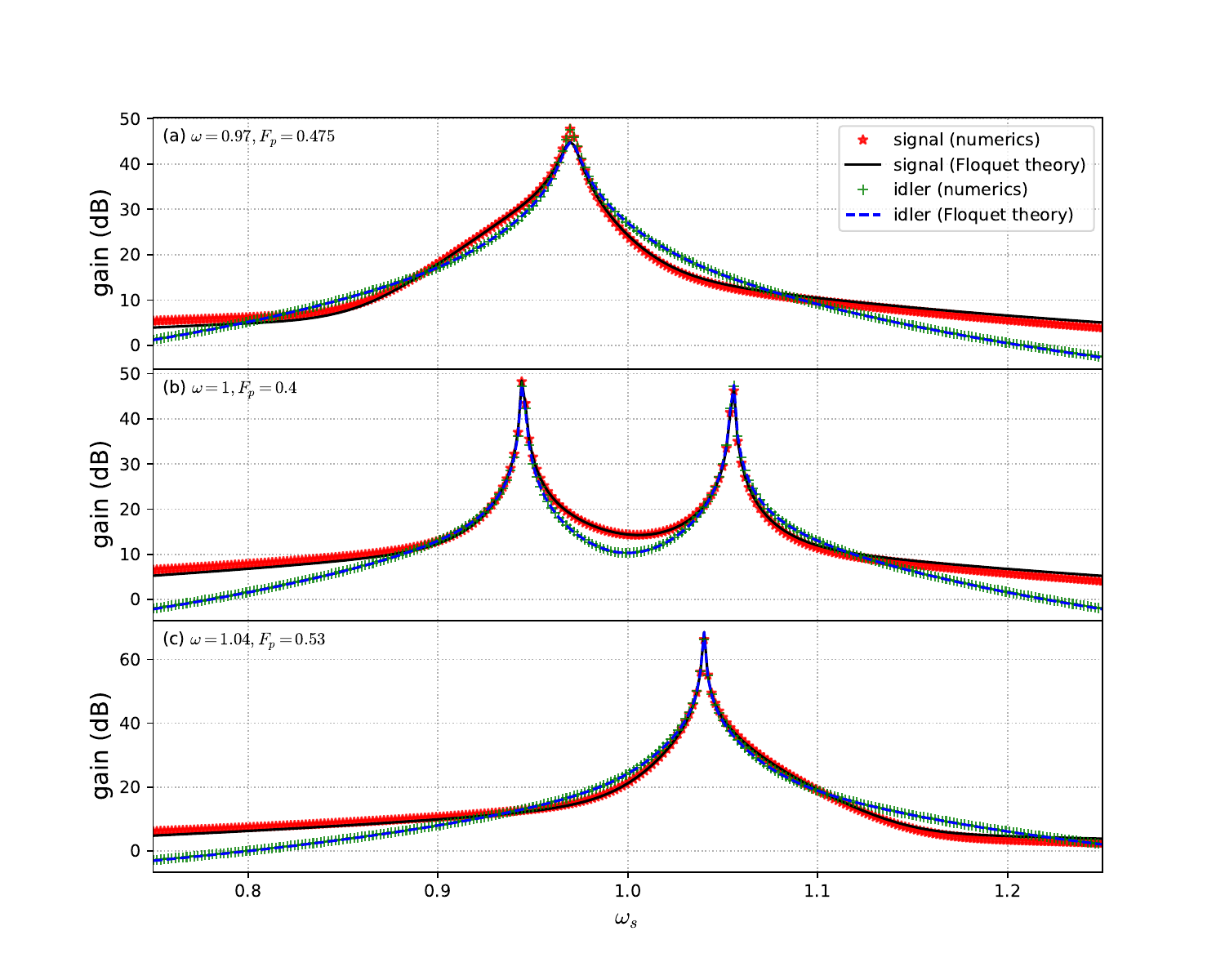}}
\caption{Coupled parametric amplifier harmonic resonator model defined in
Eq.~\eqref{2ModeParamp}. 
Comparison between numerical and Floquet theory signal and idler response gains
of the parametric amplifier.
The signal gain in dB scale is given by $20\log_{10}(r_s/F_0)$
and the idler gain in dB is given by $20\log_{10}(r_i/F_0)$.
The amplitudes $r_s$ and $r_i$ were obtained from Eq.~\eqref{signal_idler}. 
In frames a) and c) the Floquet exponents are real, whereas in frame b) they are
complex, what causes the split peaks in the signal and idler responses.  Common
parameters used and not indicated in the figure are: $\gamma_1=\gamma_2=0.1$,
$\omega_2=1.01$, $\beta_1=\beta_2=0.15$, and $\phi_0=0$.  }
\label{fig:signal_idler2}
\end{figure}

In Fig. \ref{fig:signal_idler2}, we show Floquet theory and numerical results
of the signal and idler gains, given by Eq.~\eqref{signal_idler}, as a function of $\omega_s$ for the coupled
parametric amplifier model of Eq.~\eqref{2ModeParamp}.
Again only the lowest terms of the Floquet theory expansion are
enough to give a very accurate representation of the response of the parametric
amplifier.

\begin{figure}[!ht]
    \centerline{\includegraphics[{scale=0.75}]{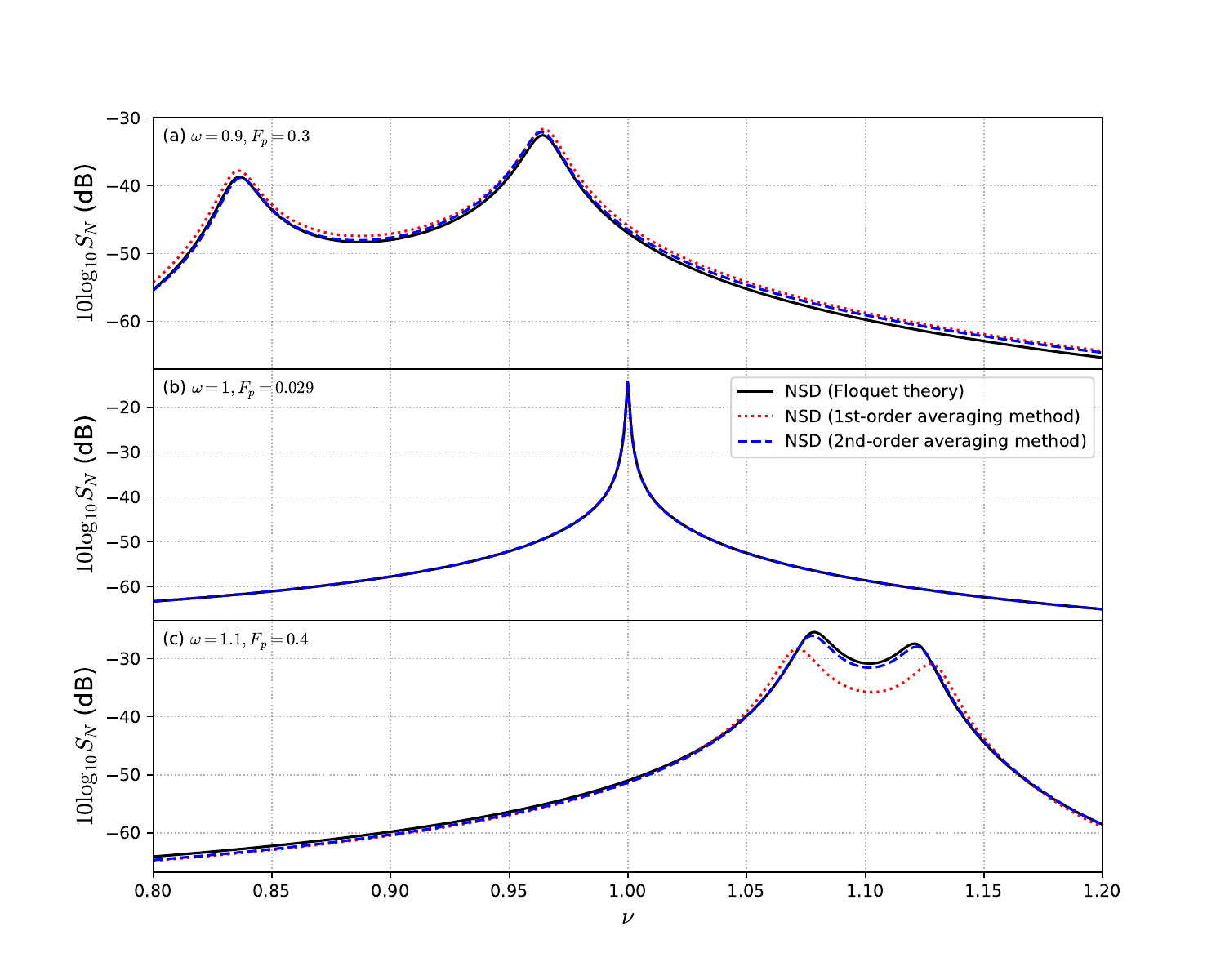}}
\caption{
The NSD $S_N$ in dB of the single parametrically-driven resonator with added
white noise.
Comparison between predictions from Floquet theory  (solid line), given by
Eq.~\eqref{S_N}, the approximate first-order averaging method (dotted
red line), and the approximate second-order averaging method (dashed blue line).
In frames a) and c), we have a parametric amplification of noise in
which the Floquet exponents are a complex-conjugate pair, whereas in frame
b), we have a parametric amplification in which the Floquet exponents are real.
In all frames $\gamma=1/65$.
}
\label{fig:NSD}
\end{figure}

\FloatBarrier
In Fig. \ref{fig:NSD}, we show the averaging method (first and second order
approximations) and the Floquet theory results for the NSD $S_N$ of the
parametric resonator with added noise given by the Eq.~\eqref{S_N}.
Here we used the noise level $D=3.08\times10^{-8}$ 
(in dimensionless units) and the quality factor $Q=65$.
These values were obtained from the electronic circuit implementation of a
parametric oscillator given in Ref.~\cite{batista2022gain}.
In this case, $x(t)$ has dimensions of voltage (measured in units of $V$) and
the noise level has dimensions of squared voltage over frequency (measured in
units of $V^2/Hz$).
To transform our dimensionless $D$ to the experimental noise level, we divide
it by $\omega_0$ the natural angular frequency of the resonator in units of
$\si{rad/s}$ and multiply it by $\si{V^2}$.
The averaging method results are obtained from the approximations to the Green's
functions Fourier transforms.
The first-order approximation to the frequency-domain response is given in Ref.~\cite{batista2022gain}.
The second-order approximation to the frequency-domain response 
is obtained from the time-domain GF given in Ref.~\cite{batista2012heating}.
In this approximation, the functional form of the GF's remains the same as in
the first-order approximation, except
that the parameters of the 2nd-order GF are obtained from 
a renormalization of the parameters of the 1st-order GF.
We clearly see that the second-order averaging method results are substantially
closer to the Floquet theory results of $S_N$.
In frame (a), first-order averaging already gives a fairly accurate result
with discrepancy of about 1 to 2 dBs, while in frame (c) first-order averaging
is off by as far as about 6dB around $\nu=1.1$.
There is a huge improvement when one goes to second-order averaging, where the
error becomes lower that 1dB at the worst case.
In frame(b), both averaging methods yield excellent agreements with the Floquet
theory results.

By obtaining the NSD by independent methods, their mutual agreement
indicates that the Floquet theory results proposed here are correct.
Further improvement in the agreement could be obtained by going to higher
orders of the averaging method, but the cost is high to justify the marginal
improvement beyond second-order approximation.
Another approach to verify our results is to use numerical integration of
stochastic differential equations.
But they are time consuming with very long time series and many integrations
(over 1000) to yield satisfactory results, especially near narrow peaks due
to resolution bandwidth limitations.
On the other hand, the Floquet theory method proposed here for calculating $S_N$
could be used as a nontrivial benchmark to test stochastic differential equation
integration methods of cyclostationary processes.
It is important to point out that Floquet theory is not a perturbative method
as the AM is.
Its range of applicability is not limited to small coefficients and near
resonance (in the case of the parametric amplifier described by
Eq.~\eqref{paramp} the parameters $\gamma$, $F_p$, and the detuning $\omega-1$
have to be small compared to 1).
Unlike the AM, in principle, Floquet theory can be accurately applied 
with any choice of parameters.
It could be used to investigate fluctuations in parametrically-driven
systems near higher parametric resonances \cite{turner1998five}. 
Another advantage of the Green's function method based on Floquet theory 
over the one based on the AM, is that it can be more easily applied to 
coupled resonators in which at least one is parametrically driven.
The same theoretical framework that we developed here could be applied,
while the AM application is usually done case by case and becomes far more
cumbersome as the dimensionality increases.

\FloatBarrier
\section{Conclusion}
\label{conclusion}
Here, we obtained the frequency-domain Green's functions of
parametrically-driven periodic dynamical systems using Floquet theory.
From this, we wrote the Fourier transform of the response of the parametric
amplifier to an added external drive.
Near the parametric instability threshold, this response can be simplified to
include only the contribution of a single real Floquet exponent that is closest
to zero or of a complex conjugate pair of Floquet exponents whose real part is closest to zero.
In this limit, the resulting expression for the response is thus considerably
simplified.
The calculated response of the parametrically-driven system to an added ac drive
or to noise, as can be seen in Eq.~\eqref{fundEq2}, is more intuitive and
simpler than equivalent ones in the literature.
We achieved this by splitting the time-domain Green's function in two pieces:
one that is translationally invariant in time, i.e. dependent only on $t-t'$ and the other that is independently
a function of $t$ and $t'$.
This simple observation led to a better understanding of the frequency-domain
behavior of the response of the parametrically excited dynamical system.
Furthermore, we also provide an expression for the NSD (Eq.~\eqref{S_N}) in
parametrically-driven systems that is very accurate and extends our previous
work \cite{batista2022gain}. 
To validate our model, we compared our results on the PSD with numerical
integration results for signal and idler gain and on the NSD with the equivalent
results based on the averaging method in the first and second order
approximations.
We obtained considerably better agreement of the excited system frequency-domain
response between the second-order approximation and the Floquet theory results.

In addition to these results, we improved upon previous models by
Wiesenfeld \etal on amplification of small signals near bifurcation points and
also on noisy precursors of bifurcations by making them more quantitative.
Although, we did not investigate the effect of the several types of bifurcations
on amplification, the results presented here are general enough to be able to
apply to all of them.
Furthermore, we believe that our contribution may help the development of
amplifiers and sensors near bifurcation points such as parametrically-driven
nano-mechanical resonators \cite{mahboob2014two}, MEMS \cite{pribovsek2022parametric},
and Josephson parametric amplifiers \cite{yamamoto2008flux, mahboob2022three}.
It could also be used in the stability analysis of limit cycles in multimode
nonlinear resonators \cite{kecskekler2023multimode}.

We believe that in addition to the small signal amplification application, the
theory developed here is also relevant to explaining the generation of frequency
combs in parametrically-driven nonlinear dynamical systems
\cite{batista2020frequency}.
In several of these systems, the frequency-comb spectrum was experimentally
observed to occur near bifurcation points
\cite{ganesan2018phononic,czaplewski2018bifurcation}.
Hence, a better understanding of the linear behavior of these dynamical systems
provided by the Green's function method based on Floquet theory that we
developed here may be relevant in guiding
future developments in this field.
This method may also be extended to investigate the linear response of limit
cycles in nonlinear dynamical systems to added small ac perturbations or
noise.
In this way, one could calculate the amplification gain near several types of
bifurcation points.
\section*{Appendix}
\label{FT}
For completeness, we present Floquet's theorem adapted from Ref.~\cite{verh96}.
\begin{thm*}[Floquet's theorem]
Given the linear ODE system
\beq
\dot x=A(t)x,
\label{non-autonomous_system}
\eeq
in which $A(t)$ is a continuous periodic $n\times n$ matrix such that $A(t+T)=A(t)$. 
The fundamental matrix $\Phi(t)$ of this ODE system can be written as
the matrix product
\begin{eqnarray}
\Phi (t)=P(t)e^{Bt},
\label{floq2}
\end{eqnarray}
in which $P(t+T)=P(t)$ is periodic and  $B$ is a constant $n\times n$ matrix.
\end{thm*}

\begin{proof}
Since the Eq.~\eqref{non-autonomous_system} has a unique solution for any initial
conditions, the fundamental matrix has an inverse.
By differentiating the operator $\Phi(t)\Phi^{-1}(t)=\mathbb{1}$, we obtain 
\begin{eqnarray}
    \frac{d\Phi^{-1}}{dt}=-\Phi^{-1}\frac{d\Phi}{dt} \Phi^{-1}=-\Phi^{-1} A(t)\Phi \Phi^{-1}=-\Phi^{-1} A(t).
\label{floq3}
\end{eqnarray}
Differentiating  the operator $\Phi^{-1}(t)\Phi(t+T)$, we find 
\begin{eqnarray}
\frac{d}{dt}[\Phi^{-1}(t)\Phi(t+T)] &=& -\Phi^{-1}(t) A(t) \Phi(t+T)+\Phi^{-1}(t) A(t+T) \Phi(t+T) \nonumber \\
&=& -\Phi^{-1}(t) A(t) \Phi(t+T)+\Phi^{-1}(t) A(t) \Phi(t+T)=0, \nonumber
\label{floq4}
\end{eqnarray}
hence $\Phi^{-1}(t)\Phi(t+T)$ is a constant in time.
This can be written as $\Phi^{-1}(t)\Phi(t+T)=\Phi(T)=C$, since
$\Phi(0)=\mathbb{1}$.
This implies that $\Phi(t+T)=\Phi(t)C$.

As we can always find a matrix $B$ such that $C=e^{BT}$ (see Lemma 7.1 of
Ref.~\cite{hale1969ordinary}), we find that
\[
\Phi(t+T)e^{-B(t+T)}=\Phi(t)e^{BT}e^{-B(t+T)}=\Phi(t)e^{-Bt}.
\]
This means that $P(t)\doteq \Phi(t)e^{-Bt}=P(t+T)$ is periodic.
Hence, $\Phi(t)=P(t)e^{Bt}$.
\end{proof}

%
\end{document}